\documentclass[aps,prl,reprint,groupedaddress,superscriptaddress]{revtex4-2}

\usepackage{graphicx} 
\usepackage{amsmath} 
\usepackage[colorlinks=true,allcolors=blue]{hyperref} 
\usepackage[capitalise]{cleveref} 
\usepackage{soul} 

\begin{document}

\title{Fully anisotropic superconductivity with few Helmholtz Fermi-surface harmonics}

\author{Jon Lafuente-Bartolome}
\author{Idoia G. Gurtubay}
\author{Asier Eiguren}
\affiliation{Materia Kondentsatuaren Fisika Saila, University of the Basque Country UPV/EHU, 48080 Bilbao, Basque Country, Spain.}
\affiliation{Donostia International Physics Center (DIPC), Paseo Manuel de Lardizabal 4, 20018 Donostia-San Sebasti\'{a}n, Spain}

\date{\today}

\begin{abstract}
We present an alternative representation for the anisotropic Eliashberg equations of superconductivity,
whose numerical solution yields an efficiency gain of several orders of magnitude with respect to the conventional representation in momentum space.
Our method is a practical realization of a long-sought approach,
whose essence is a linear transformation from regular \textbf{k} space to a set of orthonormal functions
defined as the solutions of the Helmholtz equation on the Fermi surface.
In this way, all the anisotropy of the problem can be described by a handful of coefficients with built-in symmetry.
We perform benchmark calculations on the gap anisotropy of MgB$_2$,
and reproduce previous results at a remarkably reduced computational cost.
Furthermore, we apply our methodology to efficiently determine the transition temperature of the compressed YH$_6$ hydride,
obtaining very good agreement with recent experimental measurements.
The simplification introduced by our method enables the high-throughput exploration of superconducting materials without having to resort to the isotropic approximation,
and opens up possibilities towards first principles calculations of more advanced theories of superconductivity.
\end{abstract}


\maketitle


The microscopic theory of superconductivity put forward by Bardeen, Cooper and Schrieffer \cite{BCS1957} stands for one of the greatest achievements of condensed matter theory,
as it provided the first quantitative explanation of the different experimental signatures of superconductivity available at the time.
The frequency-dependence of the superconducting gap found soon after in strong-coupling superconductors \cite{GiaeverPR1962},
was successfully rationalized by the extension of the theory developed by Eliashberg \cite{EliashbergJETP1960,ScalapinoPR1966}, which accounted for retardation effects in the electron-phonon interaction.
The discovery of superconductivity in MgB$_2$ at $39~\mathrm{K}$ \cite{NagamatsuNAT2001}
and its multiple-gap structure \cite{BouquetPRL2001,SzaboPRL2001,TsudaPRL2001,GiubileoPRL2001}
challenged the theory once again, as it added another crucial aspect to consider:
the anisotropy of the electron-phonon interaction \cite{ChoiNAT2002}.
The development of numerical methods to compute electron-phonon interactions from first principles has witnessed an enormous progress thereafter \cite{GiustinoRMP2016},
and a detailed theoretical account of experimentally measured anisotropic superconducting properties is possible nowadays  \cite{MarginePRB2013,HeilPRL2017,KawamuraPRB2017,Boeri2020}.

The advent of high-temperature superconductivity in hydrides at high pressures has resulted in a change of paradigm in superconductivity research,
in which experimental efforts are guided by prior theoretical predictions \cite{FloresLivasPHR2020,PickardARCMP2020}.
This synergy has led to the discovery of the superconductors with the highest critical temperature up to date \cite{DrozdovNAT2015,DrozdovNAT2019,SomayazuluPRL2019}.
Advanced structure searching algorithms are constantly expanding the range of possible candidates \cite{SunPRL2019},
but due to the exceedingly high computational burden associated with a full account of the anisotropy,
predictions on the critical temperature almost invariably assume an isotropic electron-phonon interaction,
and in most cases are based on the semi-empirical McMillan-Allen-Dynes formula \cite{McMillanPR1968,*AllenDynesPRB1975}.
The urgency to include full anisotropic resolution in the systematic predictions of superconducting properties in the vast range of possible interesting candidates asks for further methodological developments.

A particularly elegant and promising scheme in this direction was proposed by Allen \cite{AllenPRB1975}.
By rewriting the electron self-energy in terms of an orthonormal set of functions, the so-called Fermi-surface harmonics \mbox{(FSH),}
he showed that the anisotropic Eliashberg equations of superconductivity could take a particularly simple form \cite{AllenMitrovic1983}.
The key advantage comes from replacing the continuous integrals in \textbf{k} space by discrete sums in FSH coefficients,
where one can apply a cutoff and reduce the size of the problem dramatically without losing accuracy,
provided that those sums converge rapidly.
However, the technical difficulties to implement the specific basis set proposed in Ref.~\cite{AllenPRB1975}
has turned the practical realization of the original idea unattainable.

In this work, we present a reformulation of the Eliashberg equations in terms of an alternative basis set,
composed of the solutions of the Helmholtz equation defined on the Fermi surface, namely, the Helmholtz Fermi-surface harmonics (HFSH) \cite{EigurenNJP2014}.
We explicitly show that this representation turns out to be strikingly beneficial in the problem of superconductivity,
reducing the computational workload in several orders of magnitude.
The robustness of the numerical procedure to obtain the HFSH functions allows for a systematic application of the method in diverse materials with different crystal structures or Fermi surface topologies.
Additional improvements in the method \cite{AccompanyingPaper} provide a proper account of the gap symmetry,
and at the same time reduce the size in the expansions even further.
We perform benchmark calculations in the paradigmatic anisotropic superconductor MgB$_2$,
and determine the critical temperature of the recently synthesized YH$_6$ under pressure within full anisotropic accuracy with a handful of coefficients.


\begin{figure*}[ht]
 \includegraphics[width=2.0\columnwidth]{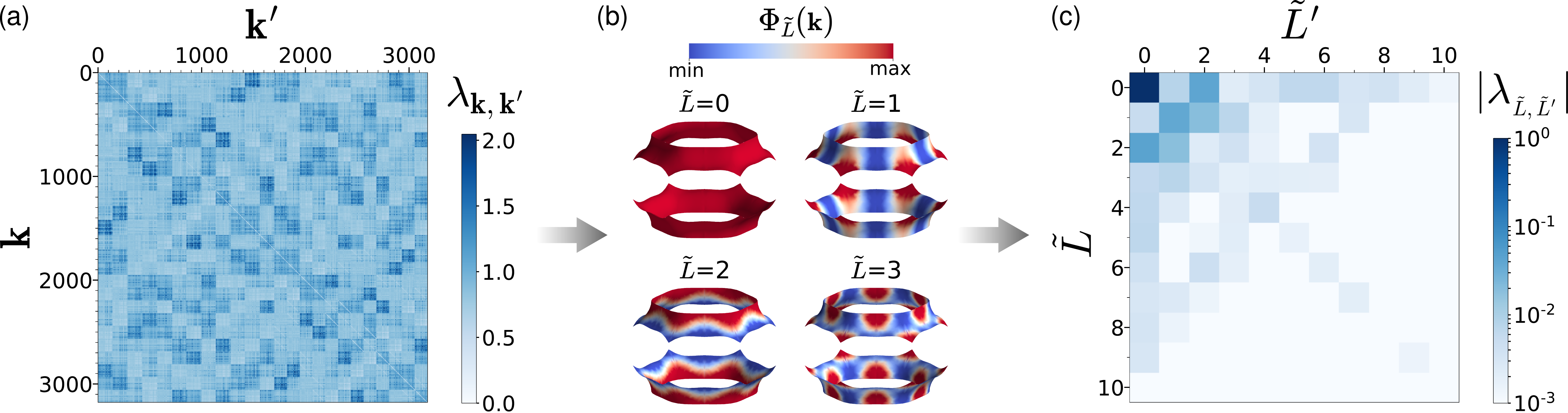}
 \caption{(a) Two-index electron-phonon mass enhancement parameter $\lambda_{{\bf k},{\bf k}'}\equiv\lambda_{{\bf k},{\bf k}'}(i\omega=0)$ computed from first principles on a discretized mesh of triangular vertices on the outer $\sigma$ Fermi surface sheet of MgB$_2$, unfolded into a matrix representation. In this example, the isosurface is formed by $\sim 3 \times 10^{3}$ vertices.
 (b) First four fully symmetric Helmholtz Fermi-surface harmonics (HFSH) basis functions on this Fermi surface sheet.
 (c) Magnitude, in logarithmic scale, of the first $10 \times 10$ fully symmetric HFSH coefficients of the two-index mass enhancement parameter $\lambda_{\tilde{L},\tilde{L}'}$ on this Fermi surface sheet. Coefficients for larger values of $\tilde{L}$ are smaller than $10^{-3}$ in magnitude.
   \label{fig:Fig1}}
\end{figure*}

We start by briefly reviewing the anisotropic Eliashberg theory of phonon-mediated superconductivity.
More detailed derivations and discussions can be found, for example, in Ref.~\cite{AllenMitrovic1983}.

For most metals, the characteristic phonon energies ($\omega_{D}$) are much smaller than the electronic energies ($\varepsilon_{F}$), that is $\omega_{D}/\varepsilon_{F} \ll 1$.
In this regime, the Migdal approximation \cite{MigdalJETP1958} in which the Eliashberg theory relies, remains valid.
This very same fact restricts the phonon-mediated superconducting pairing to a very narrow window around the Fermi surface.
As a result,
the problem of superconductivity is reduced to the solution of two coupled nonlinear integral equations defined on the Fermi surface \cite{AllenMitrovic1983}:
\begin{align}
  Z_{{\bf k}}(i\omega_{j}) & = 1 + \frac{\pi T}{\omega_{j} N_{F} \Omega_{\mathrm{BZ}}}
                                    \sum_{j'} \int_{S_{F}} \frac{ds_{\bf k'}}{v_{{\bf k}'}}
                                    ~ R^{Z}_{{\bf k}'}(i\omega_{j'}) \nonumber \\
                                    & \quad \times \lambda_{{\bf k},{\bf k}'}(i\omega_{j}-i\omega_{j'}) ~, \label{eq:elik1} \\[2ex]
  \phi_{{\bf k}}(i\omega_{j}) & = \frac{\pi T} {N_{F} \Omega_{\mathrm{BZ}}}
                                   \sum_{j'} \int_{S_{F}} \frac{ds_{\bf k'}}{v_{{\bf k}'}}
                                   ~ R^{\phi}_{{\bf k}'}(i\omega_{j'}) \nonumber \\
                                   & \quad \times [ \lambda_{{\bf k},{\bf k}'}(i\omega_{j}-i\omega_{j'}) - \mu^{*}(\omega_{c}) ] ~, \label{eq:elik2}
\end{align}
where band indices have been omitted for simplicity, and the following auxiliary definitions have been used:
\begin{subequations} \label{eq:aux_func_elik}
\begin{align}
  R^{Z}_{{\bf k}}(i\omega_{j}) = \frac{\omega_{j}Z_{{\bf k}}(i\omega_{j})}{ \sqrt{[\omega_{j} Z_{{\bf k}}(i\omega_{j}) ]^2 +    \phi_{{\bf k}}(i\omega_{j})^2}}~,~ \\
  R^{\phi}_{{\bf k}}(i\omega_{j}) = \frac{\phi_{{\bf k}}(i\omega_{j})}{\sqrt{[\omega_{j} Z_{{\bf k}}(i\omega_{j}) ]^2 + \phi_{{\bf k}}(i\omega_{j})^2}}~.~
\end{align}
\end{subequations}
In these expressions,
$N_{F}$ is the density of states at the Fermi surface,
$v_{{\bf k}}$ is the electron velocity and $\Omega_{\mathrm{BZ}}$ is the volume of the Brillouin zone.
The self-consistent solution of these coupled equations yields the renormalization factor $Z_{{\bf k}}(i\omega_{j})$ and the pair field $\phi_{{\bf k}}(i\omega_{j})$ at a given temperature $T$, where $\omega_{j}=(2j+1)\pi T$ are the Matsubara frequencies, $j$ being integer numbers.
Only for temperatures below the superconducting transition temperature ($T \leq T_{c}$) will the resulting pair-field $\phi$ be finite.
Following the most typical practice, the Coulomb repulsion has been approximated by the Morel-Anderson pseudopotential $\mu^{*}(\omega_{c})$ \cite{MorelPRB1962} with a cutoff frequency of the order of $\omega_{c} \sim 10 ~ \omega_{D}$.
All the anisotropy and retardation effects of the electron-phonon interaction are contained in
$\lambda_{{\bf k},{\bf k}'}(i\omega)$,

which is defined as \cite{GiustinoRMP2016},
\begin{equation} \label{eq:lambda_kkp}
  \lambda_{{\bf k},{\bf k}'}(i\omega) = N_{F} \sum_{\nu} \frac{2 \, \omega_{{\bf k'}-{\bf k},\nu}}{\omega_{{\bf k'}-{\bf k},\nu}^{2} + \omega^{2}}
                                        \, |g_{{\bf k},{\bf k'}}^{\nu}|^{2} ~,
\end{equation}
where $\omega_{{\bf k'}-{\bf k},\nu}$ is the frequency of a phonon mode $\nu$ with momentum ${\bf q}\equiv{\bf k'}-{\bf k}$,
and $g_{{\bf k},{\bf k'}}^{\nu}$ is the electron-phonon matrix elements for the scattering between states ${\bf k'}$ and ${\bf k}$ through a phonon ${\bf q}\nu$.
All the elements entering Eq.~(\ref{eq:lambda_kkp}) can be computed entirely from first principles at a reasonable cost nowadays.

Nevertheless, for cases in which $\lambda_{{\bf k},{\bf k}'}$ varies considerably within the Fermi surface,
an extremely fine sampling of \textbf{k} points is needed for a converged numerical integration
of Eqs.~(\ref{eq:elik1}) and (\ref{eq:elik2}),
making their direct self-consistent solution a challenging task.

An alternative reformulation of Eqs.~(\ref{eq:elik1})--(\ref{eq:aux_func_elik}) can be obtained by expanding
all the scalar quantities ---~denoted in general by $f_{{\bf k}}$~--- in terms of the complete and orthonormal basis set
fulfilling the Helmholtz equation on the Fermi surface $\{ \Phi_{L}({\bf k}) \}$ \cite{EigurenNJP2014},
\begin{equation} \label{eq:HFSH_expansion}
  f_{{\bf k}} = \sum_{L} ~ f_{L} ~ \Phi_{L}({\bf k}) ~,
\end{equation}
so that Eqs.~(\ref{eq:elik1}) and (\ref{eq:elik2}) take the form
\begin{align}
  Z_{L}(i\omega_{j}) = & ~ \delta_{L0} + \frac{\pi T}{\omega_{j}}
                                    \sum_{j'L'}
                                    ~ R^{Z}_{L'}(i\omega_{j'}) \nonumber \\
                                    & \times \lambda_{L,L'}(i\omega_{j}-i\omega_{j'}) ~, \label{eq:eliFSH1} \\[2ex]
  \phi_{L}(i\omega_{j}) = & ~ \pi T
                                   \sum_{j'L'}
                                   ~ R^{\phi}_{L'}(i\omega_{j'}) \nonumber \\
                                   & \times [ \lambda_{L,L'}(i\omega_{j}-i\omega_{j'}) - \mu^{*}(\omega_{c}) ~ \delta_{L0,L'0} ] ~. \label{eq:eliFSH2}\end{align}

In this HFSH representation,
all the anisotropy of the electron-phonon interaction is encoded in the coefficients,
\begin{equation} \label{eq:lambdaLLp}
  \lambda_{L,L'}(i\omega) = \frac{ \int_{S_{F}} \frac{ds_{\bf k}}{v_{{\bf k}}} ~ \int_{S_{F}} \frac{ds_{\bf k'}}{v_{{\bf k}'}} ~ \lambda_{{\bf k},{\bf k}'}(i\omega) ~ \Phi_{L}({\bf k}) ~ \Phi_{L'}({\bf k'}) }{ \int_{S_{F}} \frac{ds_{\bf k}}{v_{{\bf k}}} ~ \int_{S_{F}} \frac{ds_{\bf k'}}{v_{{\bf k}'}} } ~.
\end{equation}

If the coefficients $\lambda_{L,L'}$ are shown to decay rapidly for increasing indices,
a cutoff can be applied in the sums of Eqs.~(\ref{eq:eliFSH1}) and (\ref{eq:eliFSH2}) without any loss accuracy.
Moreover,
in the case of conventional $s$-wave superconductors, both $Z_{\bf k}$ and $\phi_{\bf k}$ must be invariant under all the symmetry operations of the crystal.
As a result, only the fully symmetric HFSH functions,
which we denote by the indices~$\tilde{L}$
and fulfill $\Phi_{\tilde{L}}(\mathcal{S}_{n}{\bf k}_{i}) = \Phi_{\tilde{L}}({\bf k}_{i})$ for all the $\mathcal{S}_{n}$ symmetry operations of the point group,
will contribute to their expansions ---~see Eq.~(\ref{eq:HFSH_expansion}).
In this way, Eqs.~(\ref{eq:eliFSH1}),(\ref{eq:eliFSH2}) can be effectively reduced to this fully symmetric subset.
The sparse character of $\lambda_{L,L'}$ reflects the selection rules imposed by symmetry,
which are exactly accounted for in this method.
This translates into an important reduction of the dimension of the problem,
and most importantly,
allows for a proper account of the symmetry of the computed quantities by construction.
All the details about our numerical implementation to incorporate the crystal symmetries in the HFSH basis set are described in Ref.~\cite{AccompanyingPaper}.


We now demonstrate the benefit of the transformation
by performing benchmark calculations in the paradigmatic anisotropic superconductor MgB$_2$,
for which a detailed account of the gap anisotropy has been already reported on multiple occasions \cite{ChoiNAT2002,MarginePRB2013}.
As an illustrative example, in Fig.~\ref{fig:Fig1}(a),
we represent the anisotropic $\lambda_{{\bf k},{\bf k}'}\equiv\lambda_{{\bf k},{\bf k}'}(i\omega=0)$ parameter on the outer $\sigma$ Fermi surface sheet of MgB$_2$ in a matrix form,
computed from first principles on a discrete mesh of ${\bf k},{\bf k'}$ points forming a triangularly tessellated Fermi surface
(see Ref.~\cite{AccompanyingPaper} for computational details).
This example represents a typical scenario where a dense sampling of $n_{k}\times n_{k'} \sim 10^{4}\times 10^{4}$ points  is needed to obtain a converged solution of Eqs.~(\ref{eq:elik1})--(\ref{eq:aux_func_elik}),
as $\lambda_{{\bf k},{\bf k}'}$ varies considerably from point to point on the Fermi surface.
In contrast, by transforming this quantity to the HFSH representation,
all of its anisotropic details can be described by a handful of coefficients.
We show the first four $\Phi_{\tilde{L}}({\bf k})$ functions of this sheet in Fig.~\ref{fig:Fig1}(b) for illustrative purposes,
and the magnitude of the first $\lambda_{\tilde{L},\tilde{L}'}$ coefficients,
as obtained by Eq.~(\ref{eq:lambdaLLp}), are given in Fig.~\ref{fig:Fig1}(c) in logarithmic scale.
All the elements beyond this $10 \times 10$ matrix are lower than $10^{-3}$ in magnitude,
and therefore give a negligible contribution to the sums in Eqs.~(\ref{eq:eliFSH1}) and (\ref{eq:eliFSH2}).
This implies that these equations can be solved in such a notably reduced subspace with virtually no loss of accuracy.

\begin{figure}[t]
 \includegraphics[width=0.9\columnwidth]{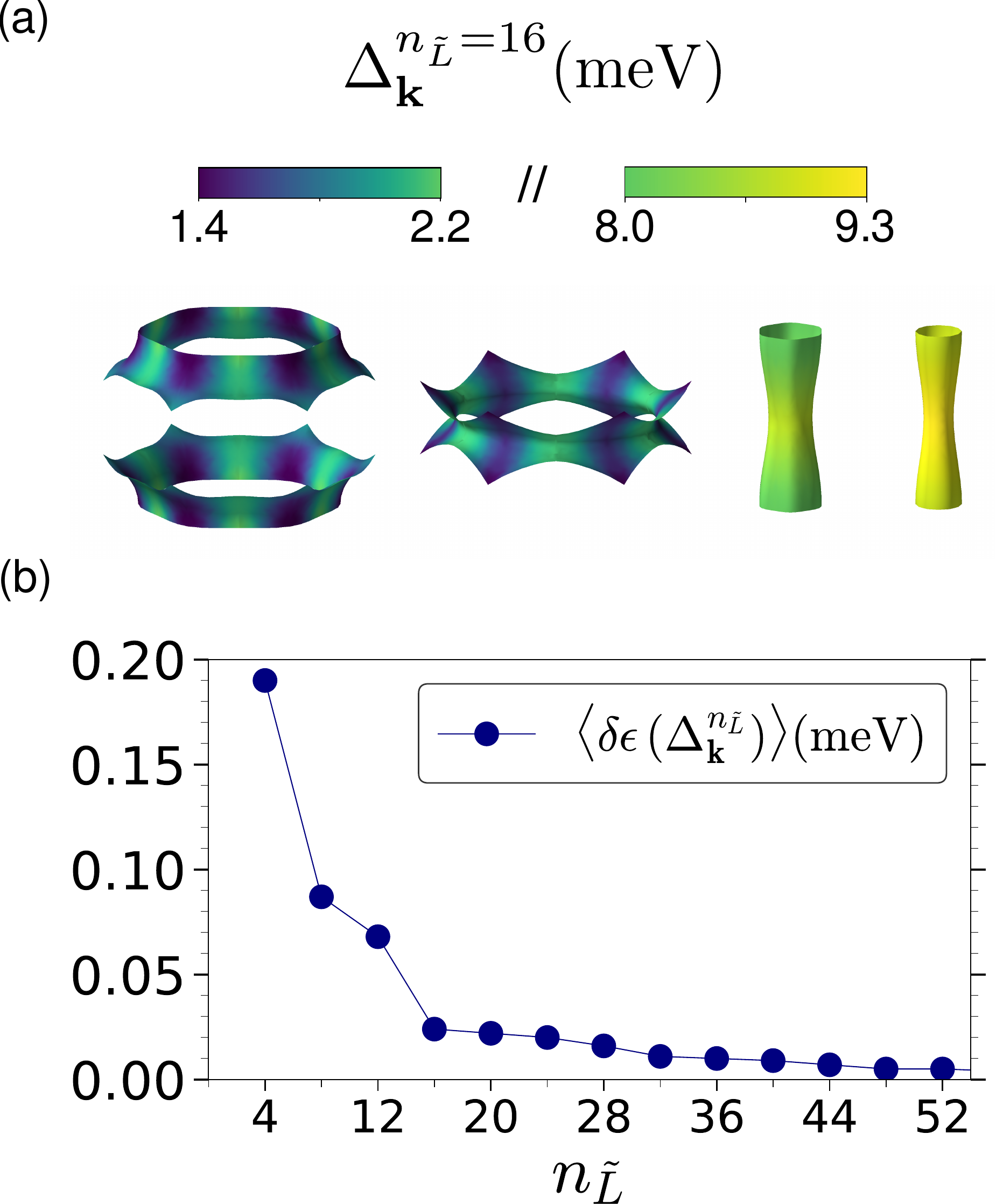}
 \caption{(a) Magnitude of the superconducting gap $\Delta_{{\bf k}}^{n_{\tilde{L}}}$  on the Fermi surface of MgB$_2$ at $10~\mathrm{K}$, obtained after solving the anisotropic Eliashberg equations in the HFSH representation, with a cutoff of $n_{\tilde{L}}=16$.
 (b) Average of the absolute error of $\Delta_{{\bf k}}^{n_{\tilde{L}}}$
 for different values $n_{\tilde{L}}$,
 with respect to the result obtained by considering all the symmetric HFSHs in the sums.
 \label{fig:Fig2}}
\end{figure}

In order to verify this assertion,
we solve Eqs.~(\ref{eq:eliFSH1}) and (\ref{eq:eliFSH2})
for MgB$_2$ at $T=10~\mathrm{K}$,
using different cutoff values in the sums, which we denote by $n_{\tilde{L}}$.
We show in Fig.~\ref{fig:Fig2} our results for the calculated superconducting gap on the Fermi surface,
\begin{equation} \label{eq:gapL2k}
  \Delta_{{\bf k}}^{n_{\tilde{L}}} =
  \frac{ \phi_{{\bf k}}^{n_{\tilde{L}}} }{ Z_{{\bf k}}^{n_{\tilde{L}}} } =
  \frac{\sum_{\tilde{L}}^{n_{\tilde{L}}} \phi_{\tilde{L}} \, \Phi_{\tilde{L}}({\bf k})}
  {\sum_{\tilde{L}}^{n_{\tilde{L}}} Z_{\tilde{L}} \, \Phi_{\tilde{L}}({\bf k})} ~,
\end{equation}
using $n_{\tilde{L}}=16$, four per Fermi surface sheet.
The Matsubara frequency cutoff has been set to ten times the maximum phonon energy,
and $\mu^{*}=0.16$ has been used.
In very good agreement with previous results \cite{MarginePRB2013}, we see that  $\Delta_{\bf k}$ clusters into two ranges of values of $(1.4,2.2)$ and $(8.0,9.3)~\mathrm{meV}$ for the $\sigma$ and $\pi$ Fermi surface sheets, respectively,
varying considerably within each sheet.

Figure ~\ref{fig:Fig2}(b) shows the average of the absolute error of $\Delta_{{\bf k}}^{n_{\tilde{L}}}$
for different values of $n_{\tilde{L}}$,
with respect to the fully converged calulation
in which all the symmetric HFSHs are considered in the sums,
\begin{equation} \label{eq:error}
  \langle \, \delta\epsilon\,(\Delta_{{\bf k}}^{n_{\tilde{L}}}) \, \rangle = \frac{\int_{S_{F}} ds_{\bf k}
                                                                         \, |\Delta_{{\bf k}}^{n_{\tilde{L}}} - \Delta_{{\bf k}}^{n_{\tilde{L}_{\mathrm{max}}}}| }
                                                                         {\int_{S_{F}} ds_{\bf k}} ~.
\end{equation}
We see that the error drops rapidly with the size of the subspace.
For a basis size as small as $n_{\tilde{L}}=16$, the error is $\sim 0.025~\mathrm{meV}$,
well below the current experimental resolution \cite{MouPRB2015}.
Besides the negligible loss of accuracy, the efficiency gain with respect to state of the art approaches is immense.
Taking Ref.~\cite{MarginePRB2013} as an example,
in order to obtain fully converged calculations for the very same system,
a Brillouin zone sampling of $n_{\bf k}=50^{3}=1.25 \times 10^{5} ~ {\bf k}$-points was needed in momentum space.
Our method, in comparison, brings an efficiency gain factor of $n_{\bf k}/n_{\tilde{L}} \sim 10^{4}$.
Another important advantage of the HFSH representation is that all the information about the superconducting state is encoded effectively in the few resulting $Z_{\tilde{L}}$ and $\Phi_{\tilde{L}}$ coefficients.
This facilitates the comparison between calculations using different meshes and the interpretation of experimental measurements, in a similar spirit as it is done when comparing Fermi surface averaged values ---~simply given by the $\tilde{L}=0$ coefficients in the HFSH representation~---,
but generalized to full anisotropic detail.

Besides the superconducting gap, one of the most important quantities characterizing a superconductor is its transition temperature $T_{c}$,
which in principle can be determined by the Eliashberg equations discussed above.
Equations (\ref{eq:elik1})--(\ref{eq:aux_func_elik}), or equivalently Eqs.~(\ref{eq:eliFSH1}) and (\ref{eq:eliFSH2}), 
can be self-consistently solved in a range of temperatures,
and the highest $T$ resulting in a non-vanishing pair amplitude $\phi$ can be identified as $T_{c}$.
However, this procedure involves several practical shortcomings.
On the one hand, in order to obtain a meaningful accuracy for the value of $T_{c}$,
the self-consistent equations have to be solved in a dense-enough range of values for $T$.
On the other hand, the nonlinear character of the equations introduces numerical difficulties to achieve self-consistency for $T\approx T_{c}$,
where the magnitude of $\phi$ becomes vanishingly small.
We have already demonstrated that the HFSH basis set remedies the first problem, as the cost of achieving self-consistency for $T \ll T_{c}$ is minimal in this representation.
In the following, we show that this basis set also provides an elegant solution to the second issue.

We start by noting that as $\phi \ll Z$ at $T \approx T_{c}$,
we can drop the $\phi^{2}$ terms in the denominators of Eq.~(\ref{eq:aux_func_elik}).
After this simplification,
Eq.~(\ref{eq:elik1}) can be inserted into Eq.~(\ref{eq:elik2}), so that we are left with a single linear equation 
for $\Delta_{\bf k}$.
This equation can be cast into an eigenvalue problem,
which after performing the transformation to the HFSH representation reads \cite{AllenMitrovic1983},
\begin{equation} \label{eq:eigvHFSH}
  \varepsilon \, \Delta_{L}(i\omega_{j}) = \sum_{j' L'} \frac{1}{|2j'+1|} \, K_{L,L'}(j,j') \, \Delta_{L'}(i\omega_{j'}) ~,
\end{equation}
where,
\begin{align} \label{eq:kernelHFSH}
  &K_{L,L'}(j,j') = \lambda_{L,L'}(i\omega_{j}-i\omega_{j'}) - \mu_{L,L'}^{*}(\omega_{c}) \nonumber \\
                              & - \delta_{jj'} \sum_{j'' L''} \Xi_{L,L'L''} ~ \lambda_{L'',0}(i\omega_{j}-i\omega_{j'})
                              ~ \mathrm{sgn}(j) ~ \mathrm{sgn}(j'') ~,
\end{align}
being $\Xi_{L,L'L''}$ the generalization of the Clebsch-Gordan coefficients for the HFSH basis set \cite{AllenPRB1975,EigurenNJP2014}.
Similar to the nonlinear equations,
the rapidly decaying values of the $\lambda_{L,L'}$ coefficients in the HFSH basis set
enable one to reduce drastically the size of the kernel $K$, and hence the dimension of the eigenvalue problem, 
while maintaining full account of the anisotropy.
The temperature at which the maximum eigenvalue $\varepsilon$ equals unity gives $T_{c}$,
since in that case the linearized Eliashberg equation is fulfilled.
The big advantage over the nonlinear equations (\ref{eq:eliFSH1}) and (\ref{eq:eliFSH2}) is that no self-consistency is needed in this case,
and that the evaluation of the auxiliary $R_{L}$ functions is not needed anymore.
%

\begin{figure}[t]
 \includegraphics[width=0.9\columnwidth]{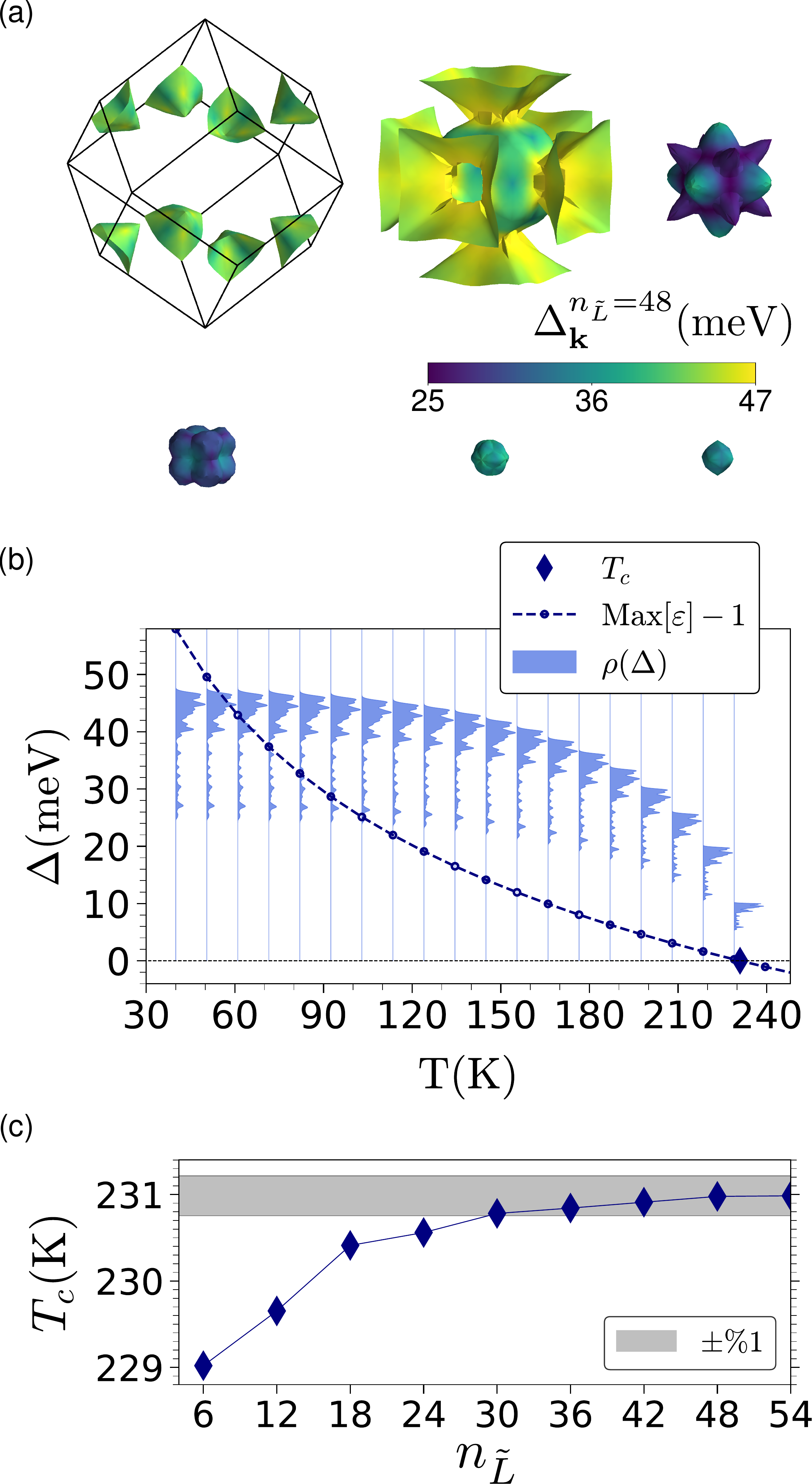}
 \caption{(a) Magnitude of the superconducting gap on the Fermi surface of YH$_6$ at $300~\mathrm{GPa}$ and $40~\mathrm{K}$.
 (b) The light blue shaded areas represent the distribution of the gap for different temperatures.
 The dark-blue dots represent the maximum eigenvalue of Eq.~(\ref{eq:eigvHFSH}) in the same range of temperatures, displaced by $-1$,
 and the dashed line is a guide to the eye.
 The temperature at which $\mathrm{Max}[\varepsilon]-1=0$ is fulfilled corresponds to $T_{c}$, and is represented by the blue diamond.
 (c) Convergence of $T_{c}$ with respect to the cutoff applied on the HFSH expansion for solving Eq.~(\ref{eq:eigvHFSH}).
 The gray shaded area represents the values within a $1\%$ accuracy with respect to the converged value, taken to be the $T_{c}$ obtained with $n_{\tilde{L}}=54$.
 \label{fig:Fig3}}
\end{figure}

We illustrate this approach using the compressed YH$_6$ hydride in its bcc structure at 300GPa as a case study (all the details of the computational setup are described in Ref.~\cite{AccompanyingPaper}).
Interestingly, the recent experimental confirmation of superconductivity in this system \cite{TroyanARXIV2019,KongARXIV2019}
has revealed a sizable deviation in the measured critical temperature with respect to the current theoretical estimates \cite{LiSCR2015,PengPRL2017,HeilPRB2019}.
For the sake of comparison, we first solved the full nonlinear Eqs.~(\ref{eq:eliFSH1}),(\ref{eq:eliFSH2}) for a set of temperatures,
where we used $\mu^{*}=0.11$ as in Ref.~\cite{HeilPRB2019}.
A reduced subspace of $n_{\tilde{L}}=48$ has been sufficient to obtain converged results.

We show our results for the superconducting gap on the six Fermi surface sheets at $40~\mathrm{K}$ in Fig.~\ref{fig:Fig3}(a).
We obtain a continuous range of values of $(25,47)$ meV for $\Delta_{\bf k}$,
being its anisotropy particularly large on the biggest sheets.
Our results are in qualitative agreement with those reported in Ref.~\cite{HeilPRB2019},
while quantitatively we obtain smaller gap values.
We trace back this discrepancy to the finer Fermi surface integrations provided by our triangulated mesh,
which also reflects in a smaller magnitude of the electron-phonon mass-enhancement parameter \cite{AccompanyingPaper}.
The distribution of the gap, $\rho(\Delta)$, obtained for different temperatures is represented by the light blue shaded areas in Fig.~\ref{fig:Fig3}(b).
The magnitude of the gap decreases with temperature,
and we do not find superconductivity ($\phi \neq 0$) beyond $\sim 230~\mathrm{K}$.

The maximum eigenvalue obtained after diagonalizing Eq.~(\ref{eq:eigvHFSH}) for the same range of temperatures and subspace size is represented by the blue dots in Fig.~\ref{fig:Fig3}(b), displaced by $-1$ for ease of visualization.
Its change with temperature is very smooth, allowing for an efficient use of root finding algorithms to detect the exact point where $\varepsilon=1$ is fulfilled.
We find $T_{c}=230.98 ~ \mathrm{K}$, in really good agreement with very recent experimental results \cite{TroyanARXIV2019,KongARXIV2019}.
With the aim of reducing the size of the problem as much as possible,
we analyze in Fig.~\ref{fig:Fig3}(c) the sensitivity of the predicted $T_{c}$ with respect to the HFSH expansion cutoff $n_{\tilde{L}}$.
Interestingly, we verify that convergence is reached very rapidly,
obtaining results within $1\%$ of accuracy with as few as 30 HFSHs.
This result demonstrates that the HFSH basis set appears extremely beneficial for a precise determination of $T_{c}$ with a full inclusion of the anisotropy,
as the problem is reduced to a small matrix diagonalization for the finite range of temperatures involved in the root finding procedure.


In conclusion, we have presented an efficient numerical scheme to predict superconducting properties from first principles with full account of the electron-phonon anisotropy.
We have shown that our method introduces a reduction of several orders of magnitude in the computational workload as compared to the conventional approach, while carrying practically no loss of accuracy.
Furthermore, we have demonstrated that our procedure is robust and generally valid for diverse systems, making it readily applicable to the high-throughput exploration of novel superconductors.
More generally,
the remarkable simplification introduced by our scheme opens the way towards new \textit{ab initio} and model theoretical treatments
since only a few coefficients are sufficient to describe the complexity of the Fermi surface,
and even the selection rules are naturally incorporated by construction.


\begin{acknowledgments}
The authors acknowledge the Department of Education, Universities and Research of the Basque Government
and the University of the Basque Country UPV/EHU (Grant No. IT756-13),
the Spanish Ministry of Economy and Competitiveness MINECO (Grants No. FIS2016-75862-P and No. PID2019-103910GB-I00)
and the University of the Basque Country UPV/EHU (Grant No. GIU18/138) for financial support.
J.L.-B. acknowledges the University of the Basque Country UPV/EHU (Grant No. PIF/UPV/16/240)
and the Donostia International Physics Center (DIPC) for financial support.
Computer facilities were provided by the DIPC.
\end{acknowledgments}

\bibliography{bibliography}

\end{document}